\documentclass[prd,twocolumn,nofootinbib]{revtex4}
\usepackage{graphicx, epsfig}
\usepackage{color}
\usepackage{mathrsfs}
\usepackage{bm}
\usepackage{amsmath,amssymb,empheq}
\usepackage{mathrsfs}
\usepackage[caption=false]{subfig}
\usepackage{hyperref}
\usepackage[normalem]{ulem}
\usepackage{mathtools}
\usepackage[x11names]{xcolor}

\definecolor{fashionfuchsia}{rgb}{0.96, 0.0, 0.63}
\colorlet{no_so_fashion_purple}{blue!50!red}

\newcommand{\be}{\begin{equation}}
\newcommand{\ee}{\end{equation}}
\newcommand{\ba}{\begin{eqnarray}}
\newcommand{\ea}{\end{eqnarray}}

\newcommand{\nn}{\nonumber}

\newcommand{\calA}{\mathscr{A}}

\def\itie{{\it i.e.}\,}

\newcommand{\calF}{\mathscr{F}}
\newcommand{\calK}{\mathcal{K}}
\newcommand{\Dpsi}{\mathcal{D}}
\newcommand{\tr}{{\rm tr}\,}
\newcommand{\tl}{_{\rm tl}}

\begin{document}
\title{8-type electric flux tube solutions in SU(3) gauge theory}
\author{Jude Pereira\footnote{jperei10@asu.edu}}
\affiliation{
Physics Department, Arizona State University, Tempe,  Arizona 85287, USA.
}

\begin{abstract}
We consider electric flux tube solutions in SU(3) gauge theory with scalar fields in the
fundamental representation, for an electric field in the $\lambda_8$ direction (8-type).
In temporal gauge, the 3-type electric field rotates within a single SU(2) subgroup at one
frequency, whereas the 8-type electric field necessarily rotates within three SU(2) subgroups
with at least two different frequencies. We construct the homogeneous 8-type electric field with
three scalar fields and show that two scalar fields are sufficient only on a one-parameter family
of gauge field backgrounds. Like the 3-type field, the homogeneous 8-type field is immune to decay
through Schwinger pair production of gauge bosons. We then construct an 8-type electric string solution in
temporal gauge.
Its electric field is of 8-type on the string axis and, unlike the 3-type string, the string
is only net color neutral, which results in a weak radial electric field.
\end{abstract}

\maketitle

\section{Introduction}
\label{intro}

Magnetic-type solutions in gauge theories, such as vortex lines~\cite{Nielsen:1973cs}, have been
studied for many decades, but electric-type solutions in non-Abelian gauge theories have been
discovered only recently. They are built on the observation by Brown and Weisberger
(BW)~\cite{BROWN1979285} that there are an
infinite number of gauge inequivalent non-Abelian gauge fields, in addition to embedded
Maxwellian gauge fields, all of which lead to identical electric fields. Electric fields resulting from
Maxwellian gauge fields are unstable to Schwinger pair production~\cite{Schwinger:1951nm} of gauge
particles and will eventually dissipate~\cite{Cardona:2021ovn}. On the other hand, electric fields
generated from the BW gauge fields need volume-filling sources but are classically stable for a
wide range of parameters~\cite{Pereira:2024sic} and also quantumly stable
(``unexciting'') to Schwinger pair production of gauge particles~\cite{Vachaspati:2022ktr}.
Electric string solutions, equivalently electric flux tubes, were first constructed in an SU(2)
gauge theory with a scalar field in the fundamental representation as the
source~\cite{Vachaspati:2022gco,Vachaspati:2023tpt}.

In a previous paper~\cite{Pereira:2025su3} the construction was extended to SU(3) gauge theory,
because of its relevance to quantum chromodynamics (QCD), where chromoelectric flux tubes between
static color sources are seen in lattice simulations~\cite{Baker:2022cwb}. Since SU(3) has two elements in its
Cartan subalgebra, there are two different ``electric fields'', one in the $\lambda^3$ direction
and the other in the $\lambda^8$ direction, and correspondingly two types of electric solutions
that we called ``3-type'' and ``8-type'' solutions. The two types cannot be related by a gauge
transformation. This can be seen from the eigenvalues of the electric field matrix
$E=E^a\lambda^a/2$, which are proportional to $(1,-1,0)$ for the 3-type and to $(1,1,-2)$ for the
8-type. Equivalently, a constant electric field in SU(3) has two independent Casimir invariants,
$E^aE^a$ and $d^{abc}E^aE^bE^c$~\cite{Nayak:2005casimir}, and the second one vanishes for the 3-type.
We will use the ratio
\be
\tau \equiv \frac{\sqrt{6}\,\tr E^3}{(\tr E^2)^{3/2}} = \frac{\sqrt{3}\,d^{abc}E^aE^bE^c}{(E^aE^a)^{3/2}}
\label{taudef}
\ee
to quantify the type of an electric field. It vanishes for the 3-type and satisfies $|\tau|=1$ for
the 8-type, while a general electric field has $0<|\tau|<1$.

In Ref.~\cite{Pereira:2025su3} we constructed the uniform 3-type electric field and the 3-type
electric string with two scalar fields in the fundamental representation, and showed that they
are immune to decay through Schwinger pair production of gauge bosons. However, we were not able
to construct an 8-type solution, and the construction of 8-type electric strings remained an open
question. In this paper we construct the 8-type solutions, and show that the homogeneous 8-type
electric field is also immune to Schwinger pair production of gauge bosons.

The construction of the 3-type solutions in Ref.~\cite{Pereira:2025su3} largely follows the
procedures developed for the SU(2) case~\cite{Vachaspati:2022gco,Vachaspati:2023tpt}. In
temporal gauge the 3-type gauge field rotates within an SU(2) subgroup of SU(3) at a single
frequency $\Omega$, and with a suitable choice of scalar fields the explicit time dependence
drops out of the equations of motion. We find that the 8-type gauge field in temporal gauge
instead rotates within three SU(2) subgroups at the frequencies $\Omega/2$, $\Omega/2$ and
$\Omega$. We also show that this is unavoidable, since a gauge field that rotates at a single
frequency in temporal gauge always gives a 3-type electric field. As a result, the homogeneous
8-type electric field generically requires three scalar fields instead of two, and the 8-type
electric string requires two different profile functions for the gauge fields, one for each
frequency. The two profile functions behave differently away from the string axis, and so the
electric field of the string is of 8-type on the axis but not everywhere. Another difference
from the 3-type string is that the 8-type string is net color neutral but not color neutral at
every point, which leaves a weak radial electric field.

We start out in Sec.~\ref{gaugefields} by describing the 8-type BW gauge fields. In
Sec.~\ref{seclambda8} we write the homogeneous 8-type electric field in temporal gauge, construct
the scalar fields that source it, and construct the electric string solution, which we summarize
in Sec.~\ref{summary}. In Sec.~\ref{unexciting} we show that the homogeneous 8-type electric field
is stable to Schwinger pair production of gauge bosons and discuss the electric string. We discuss
questions for future work in the concluding Sec.~\ref{conclusions}. In Appendix~\ref{appA} we describe a general method to source uniform BW
electric fields with scalar fields, in Appendix~\ref{appB} we prove our results on the frequencies
and the type of the electric field, in Appendix~\ref{appC} we list the full set of radial
equations for the electric string, and in Appendix~\ref{appD} we describe our numerical methods
and checks.

\section{Gauge fields}
\label{gaugefields}

As in Ref.~\cite{Pereira:2025su3}, we consider SU(3) gauge fields coupled to $N$ scalar fields in
the fundamental representation, with the Lagrangian
\be
L = -\frac{1}{4}W^{a}_{\mu\nu}W^{\mu\nu a} + \sum_{i=1}^{N}|D_{\mu}\Phi_i|^2 - V
\label{lag}
\ee
where
\be
D_{\mu}\Phi_i = \partial_{\mu}\Phi_i - i\frac{g}{2}W_{\mu}^{a}\lambda^{a}\Phi_{i} ,
\label{covDev}
\ee
\be
W^{a}_{\mu\nu} = \partial_{\mu}W^{a}_{\nu} - \partial_{\nu}W^{a}_{\mu} + g f^{abc} W^{b}_{\mu}W^{c}_{\nu} ,
\label{fieldstrength}
\ee
$f^{abc}$ are the SU(3) structure constants defined by
$[\lambda_a,\lambda_b] = 2if^{abc}\lambda_c$, and
\be
V = \sum_{i=1}^{N}\big(m_i^2|\Phi_i|^2 + \kappa_i|\Phi_i|^4\big) .
\label{potentialV}
\ee
We use the metric signature $(+,-,-,-)$. Unlike Ref.~\cite{Pereira:2025su3}, we have not
included the interaction terms $\kappa_{ij}[(\Phi_i^\dagger\Phi_j)^2+{\rm h.c.}]$ in the
potential. These terms vanished for the orthogonal scalar fields of the 3-type solution, but the
scalar fields of the 8-type solutions will not be orthogonal (see Sec.~\ref{type8scalar}).
It will often be convenient to use the matrix-valued gauge field
$\calA_\mu \equiv (g/2)W^a_\mu\lambda^a$, in terms of which the field strength is
\be
\calF_{\mu\nu} \equiv \frac{g}{2}W^a_{\mu\nu}\lambda^a
= \partial_\mu\calA_\nu - \partial_\nu\calA_\mu - i[\calA_\mu,\calA_\nu] .
\ee

A general set of BW gauge fields that result in uniform electric fields along the $z$-direction
and in the $a=3$ and $a=8$ directions in group space can be written as~\cite{Pereira:2025su3}
\be\label{BWfields}
\begin{aligned}
  W^{1}_{\mu} &= \frac{\Omega_1}{g}\partial_{\mu}t, \quad W^{2}_{\mu} = -\frac{\Omega_1 E}{\Omega^2}\partial_{\mu}z, \quad W^{4}_{\mu} = \frac{\Omega_2}{g}\partial_{\mu}t, \\
W^{5}_{\mu} &= \frac{\Omega_2 E}{\Omega^2}\partial_{\mu}z, \quad W^{6}_{\mu} = \frac{\Omega_3}{g}\partial_{\mu}t, \quad W^{7}_{\mu} = - \frac{\Omega_3 E}{\Omega^2}\partial_{\mu}z
\end{aligned}
\ee
where $\Omega^2 \equiv \Omega_1^2+\Omega_2^2+\Omega_3^2$. The non-vanishing field strength
components are
\ba
W^{3}_{\mu\nu} &=& \frac{E}{2\Omega^2}(\Omega_2^2 + \Omega_3^2 - 2\Omega_1^2)(\partial_{\mu}t\partial_{\nu}z - \partial_{\mu}z\partial_{\nu}t)
\label{W3munu}
\\
W^{8}_{\mu\nu} &=& \frac{\sqrt{3}E}{2\Omega^2}(\Omega_2^2 - \Omega_3^2)(\partial_{\mu}t\partial_{\nu}z - \partial_{\mu}z\partial_{\nu}t)
\label{W8munu}
\ea
Setting $\Omega_1^2 = (\Omega_2^2 + \Omega_3^2)/2$ results in a constant electric field in the
$a=8$ direction (8-type). Weyl-permuted versions of this condition also give 8-type electric
fields, while generic values of the $\Omega_i$ give electric fields of intermediate type. We
parametrize the 8-type gauge fields by
\be
\Omega_2 = \sqrt{2}\,\Omega_1\cos\theta, \quad \Omega_3 = \sqrt{2}\,\Omega_1\sin\theta,
\label{thetadef}
\ee
with $0\le\theta<\pi/4$. Then $\Omega^2 = 3\Omega_1^2$, $W^3_{\mu\nu}=0$ and
\be
W^{8}_{\mu\nu} = \frac{E}{\sqrt{3}}\cos 2\theta\, (\partial_{\mu}t\partial_{\nu}z - \partial_{\mu}z\partial_{\nu}t) .
\ee
For most of this paper we will set $\theta=0$, \itie $\Omega_3=0$ and $\Omega_2=\sqrt{2}\,\Omega_1$,
which is the 8-type analog of the choice $\Omega_2=\Omega_3=0$ made for the 3-type electric field
in Ref.~\cite{Pereira:2025su3}. This results in the gauge fields
\ba
W^{1}_{\mu} &=& \frac{\Omega}{\sqrt{3}\,g}\partial_{\mu}t, \quad W^{2}_{\mu} = -\frac{E}{\sqrt{3}\,\Omega}\partial_{\mu}z, \nn \\
W^{4}_{\mu} &=& \frac{\sqrt{2}\,\Omega}{\sqrt{3}\,g}\partial_{\mu}t, \quad W^{5}_{\mu} = \frac{\sqrt{2}\,E}{\sqrt{3}\,\Omega}\partial_{\mu}z ,
\label{BWslice}
\ea
with all other $W^b_\mu=0$. The only non-vanishing field strength component is
\be
W^{8}_{\mu\nu} = \frac{E}{\sqrt{3}}(\partial_{\mu}t\partial_{\nu}z - \partial_{\mu}z\partial_{\nu}t) .
\ee
As in Ref.~\cite{Pereira:2025su3} we define
\be
\epsilon = \frac{gE}{\Omega} .
\ee
Note that the same value of $\epsilon$ does not correspond to the same electric field for the two
types. For the 3-type field the magnitude of the electric field is $E$ and the energy density is
$E^2/2$, whereas for the 8-type field in \eqref{BWslice} they are $E/\sqrt{3}$ and $E^2/6$.

The gauge four-currents are defined as
\be\label{fourcurrent}
j^{\mu a} \equiv D_{\nu}W^{\mu\nu a} = \partial_{\nu}W^{\mu\nu a} + g f^{abc} W^{b}_{\nu}W^{\mu\nu c}
\ee
and are given in Eqs.~(10)--(15) of Ref.~\cite{Pereira:2025su3} for the general BW gauge fields
\eqref{BWfields}. For the gauge fields in \eqref{BWslice} the only non-vanishing components are
\be
j^{4}_{\mu} = -\frac{\epsilon^2\Omega}{\sqrt{6}\,g}\partial_{\mu}t , \quad
j^{5}_{\mu} = \frac{\epsilon\Omega^2}{\sqrt{6}\,g}\partial_{\mu}z .
\label{BWcurrents}
\ee

In what follows, we will investigate 8-type electric fields and construct electric string
solutions following Refs.~\cite{Vachaspati:2023tpt,Pereira:2025su3}.

\section{Electric field in $\lambda_8$ direction}
\label{seclambda8}

\subsection{Homogeneous electric fields}
\label{homogeneousE}

The time components of the gauge fields in \eqref{BWslice} can be written in matrix form as
$\calA_t = H$ with
\be
H = \frac{\Omega}{2\sqrt{3}}\big(\lambda^1+\sqrt{2}\lambda^4\big)
= \frac{\Omega}{2}\big(u_+u_+^T - u_-u_-^T\big)
\label{Hmatrix}
\ee
where $u_\pm = (e_1\pm n)/\sqrt{2}$, $e_1=(1,0,0)^T$ and $n=(0,1,\sqrt{2})^T/\sqrt{3}$. The third
eigenvector of $H$, $u_0 = (0,\sqrt{2},-1)^T/\sqrt{3}$, has eigenvalue zero. We can write the gauge
fields in the temporal gauge $W^a_t=0$ by acting with the gauge transformation
\be
W_{\mu} \to W'_{\mu} = U W_{\mu}U^{\dagger} + \frac{2i}{g}U\partial_{\mu}U^{\dagger}
\label{gaugetransform}
\ee
where $W_\mu = W^a_\mu\lambda_a$, as in Ref.~\cite{Pereira:2025su3}. It can be checked that
\be
U = \exp\left(-i\frac{\Omega t}{2}\Lambda\right) R_0^T ,
\label{Udef}
\ee
where $\Lambda\equiv{\rm diag}(1,0,-1) = (\lambda^3+\sqrt{3}\lambda^8)/2$ and $R_0 = (u_+,u_0,u_-)$
is the orthogonal matrix with columns $u_\pm$ and $u_0$, results in temporal gauge. The matrix
$R_0$ rotates $H$ into the diagonal form $\Omega\Lambda/2$ and the exponential then removes
it.\footnote{Since $\det R_0=-1$, $U$ is not an element of SU(3). However, $-U\in{\rm SU(3)}$ acts
in the same way on the gauge fields and only changes the signs of the scalar fields, which is a
symmetry of the Lagrangian \eqref{lag}.} The gauge fields in temporal gauge are
\ba
W^{1}_{\mu} &=& \frac{2\epsilon}{3g}\sin\frac{\Omega t}{2}\,\partial_{\mu}z, \ \
W^{2}_{\mu} = -\frac{2\epsilon}{3g}\cos\frac{\Omega t}{2}\,\partial_{\mu}z, \nn \\
W^{6}_{\mu} &=& -\frac{2\epsilon}{3g}\sin\frac{\Omega t}{2}\,\partial_{\mu}z, \ \
W^{7}_{\mu} = \frac{2\epsilon}{3g}\cos\frac{\Omega t}{2}\,\partial_{\mu}z, \nn \\
W^{4}_{\mu} &=& \frac{\epsilon}{3g}\sin\Omega t\,\partial_{\mu}z, \ \
W^{5}_{\mu} = -\frac{\epsilon}{3g}\cos\Omega t\,\partial_{\mu}z,
\label{gaugeField}
\ea
along with $W^{3}_{\mu} = W^8_\mu = 0$. This should be compared with the 3-type gauge fields in
temporal gauge, $W^{1}_{\mu} = -(\epsilon/g)\cos\Omega t\,\partial_{\mu}z$ and
$W^{2}_{\mu} = -(\epsilon/g)\sin\Omega t\,\partial_{\mu}z$~\cite{Pereira:2025su3}. The 3-type gauge
field rotates within the SU(2) subgroup generated by $\lambda^1$, $\lambda^2$ and $\lambda^3$ at
the frequency $\Omega$. The 8-type gauge field \eqref{gaugeField} instead rotates simultaneously
within the three SU(2) subgroups generated by $(\lambda^1,\lambda^2)$, $(\lambda^6,\lambda^7)$ and
$(\lambda^4,\lambda^5)$, at the frequencies $\Omega/2$, $\Omega/2$ and $\Omega$ respectively.

The non-vanishing field strength components in temporal gauge are
$W^a_{\mu\nu} = \dot W^a_z(\partial_\mu t\partial_\nu z - \partial_\nu t\partial_\mu z)$ for
$a=1,2,4,5,6,7$, where overdots denote time derivatives. For example,
\be
W^{1}_{\mu\nu} = \frac{E}{3}\cos\frac{\Omega t}{2}\,(\partial_{\mu}t\partial_{\nu}z - \partial_{\nu}t\partial_{\mu}z) ,
\ee
and each of the three SU(2) subgroups carries an electric field of magnitude $E/3$. The
gauge-invariant energy density in the electric fields
\be
\mathcal{E}_{\text{elec}} = \frac{1}{2}(W^{a}_{0i})^2 = \frac{\epsilon^2\Omega^2}{6g^2} = \frac{1}{6}E^2
\ee
is time-independent.

It will be useful to write the gauge fields \eqref{gaugeField} in matrix form as
\be
\calA_z = e^{-i\nu\Lambda t}\,\calK\,e^{i\nu\Lambda t} , \quad
\calK = i\begin{pmatrix}
0 & f_1 & f_2\\
-f_1 & 0 & -f_1\\
-f_2 & f_1 & 0
\end{pmatrix} ,
\label{Kmatrix}
\ee
with $\nu = \Omega/2$, $f_1 = \epsilon/3$ and $f_2=\epsilon/6$. The amplitude $f_1$ belongs to the
two subgroups that rotate at the frequency $\nu$, and $f_2$ to the subgroup that rotates at $2\nu$.
Using \eqref{fourcurrent}, the four-currents for the gauge fields \eqref{Kmatrix} in temporal gauge
can be computed for any constant $f_1$, $f_2$ and $\nu$ as follows
\ba
j^{1}_{\mu} &=& \frac{2\nu}{g}\big[3f_1f_2\cos\nu t\,\partial_\mu t + \nu f_1 \sin\nu t\,\partial_\mu z\big]
\label{gaugecurrent1}
\\
j^{2}_{\mu} &=& \frac{2\nu}{g}\big[3f_1f_2\sin\nu t\,\partial_\mu t - \nu f_1 \cos\nu t\,\partial_\mu z\big]
\\
j^{6}_{\mu} &=& \frac{2\nu}{g}\big[3f_1f_2\cos\nu t\,\partial_\mu t - \nu f_1 \sin\nu t\,\partial_\mu z\big]
\\
j^{7}_{\mu} &=& \frac{2\nu}{g}\big[3f_1f_2\sin\nu t\,\partial_\mu t + \nu f_1 \cos\nu t\,\partial_\mu z\big]
\\
j^{4}_{\mu} &=& \frac{8\nu^2f_2}{g}\sin 2\nu t\,\partial_\mu z
\\
j^{5}_{\mu} &=& -\frac{8\nu^2f_2}{g}\cos 2\nu t\,\partial_\mu z
\\
j^{3}_{\mu} &=& \frac{1}{\sqrt{3}}\,j^{8}_{\mu} = -\frac{2\nu}{g}\big(f_1^2+2f_2^2\big)\partial_\mu t
\label{gaugecurrent3}
\ea
Unlike the 3-type case, where the only non-vanishing time component is
$j^3_t$~\cite{Pereira:2025su3}, the time components of the currents now have static components
along $\lambda^3$ and $\lambda^8$ as well as components that rotate at the frequency $\nu$.

The appearance of more than one frequency is not a consequence of our choice $\theta=0$. Let
$\calA_t=H$ and $\calA_z=B$ be constant matrices in BW gauge. In temporal gauge, $\calA_z$ becomes
$e^{-iHt}Be^{iHt}$ (up to a constant gauge rotation) and contains the frequencies $\mu_j-\mu_k$,
where $\mu_j$ are the eigenvalues of $H$. In Appendix~\ref{appB} we show that if only one
frequency is present, \itie if $[H,[H,B]]=\chi^2 B$ for some constant $\chi$, then the electric
field $-i[H,B]$ is of 3-type. An 8-type electric field requires $H$ to have three distinct
eigenvalues and $B$ to have non-vanishing matrix elements between all pairs of eigenvectors of
$H$. The electric field then rotates with the frequencies $\mu_1-\mu_2$, $\mu_2-\mu_3$ and
$\mu_1-\mu_3$, and at least two of them are different. For \eqref{Hmatrix} the eigenvalues are
$\Omega/2$, $0$ and $-\Omega/2$, which gives the frequencies $\Omega/2$, $\Omega/2$ and $\Omega$.

We will also need the type ratio \eqref{taudef} of a gauge field of the form \eqref{Kmatrix} with
arbitrary $f_1$ and $f_2$. Using $\calF_{tz} = \partial_t\calA_z$ we obtain (see
Appendix~\ref{appB})
\be
\tau = -\frac{3\sqrt{6}}{2}\,\frac{f_1^2f_2}{(f_1^2+2f_2^2)^{3/2}} .
\label{tauf}
\ee
The electric field is of 8-type, $\tau=\mp1$, exactly when $f_2=\pm|f_1|/2$, and of 3-type when
either $f_1$ or $f_2$ vanishes. In \eqref{gaugeField} we have $f_2=f_1/2$ and $\tau=-1$.

\subsection{Scalar fields}
\label{type8scalar}

Next, we would like to obtain scalar fields that can consistently source the gauge fields
in \eqref{gaugeField}. In Ref.~\cite{Pereira:2025su3} we found that a single scalar field cannot
source the 3-type gauge fields and that two orthogonal scalar fields are needed. For the 8-type
gauge fields we will need three scalar fields\footnote{We discuss solutions with two scalar fields
in Sec.~\ref{twoscalar}.}, \itie $N=3$ in \eqref{lag}. The equation of motion for the scalar
field $\Phi_i$ is
\be
D_{\mu}D^{\mu}\Phi_i + m_i^2\Phi_i + 2\kappa_i|\Phi_i|^2\Phi_i = 0
\label{Phieq}
\ee
where, in temporal gauge and with $\Phi_i=\Phi_i(t)$,
\be
D_{\mu}D^{\mu}\Phi_i = \partial_t^2\Phi_i + \frac{g^2}{4}W^a_zW^b_z\lambda^a\lambda^b\Phi_i
= \partial_t^2\Phi_i + \calA_z^2\Phi_i .
\label{covDcovDPhi}
\ee
The four-current for the scalar fields is
\be
J^{a}_{\mu} = \frac{i g}{2} \sum_{i=1}^{3}\big[\Phi_i^{\dagger}\lambda^{a}D_{\mu}\Phi_i - \text{h.c.}\big]
\label{scalarcurrent}
\ee
and the gauge field equations of motion require $j^a_\mu = J^a_\mu$.

Since the gauge fields in \eqref{Kmatrix} depend on time through the factors $e^{\mp i\nu\Lambda t}$,
we write the scalar fields as
\ba
\Phi_1 &=& \eta_1\begin{pmatrix}
x_+ e^{-i(\omega+\nu)t}\\ x_0 e^{-i\omega t}\\ x_- e^{-i(\omega-\nu)t}
\end{pmatrix}, \
\Phi_2 = \eta_1\begin{pmatrix}
-x_- e^{i(\omega-\nu)t}\\ x_0 e^{i\omega t}\\ -x_+ e^{i(\omega+\nu)t}
\end{pmatrix}, \nn\\
\Phi_3 &=& \eta_3\begin{pmatrix}
y_+ e^{-i\nu t}\\ y_0\\ -y_+ e^{i\nu t}
\end{pmatrix}
\label{phiansatz}
\ea
where $\nu=\Omega/2$ and $x_\pm$, $x_0$, $y_+$, $y_0$ are real constants that satisfy
\be
x_+^2+x_0^2+x_-^2 = 1 , \quad 2y_+^2+y_0^2 = 1 .
\label{normalization}
\ee
Each component of the scalar fields oscillates at a single frequency. The field $\Phi_2$ is the
mirror image of $\Phi_1$ under the reflection $\sigma={\rm diag}(-1,1,1)$ in the original basis
combined with $(t,z)\to(-t,-z)$, while $\Phi_3$ is mapped to itself. The pair $\Phi_1$, $\Phi_2$
plays the role of the pair of scalar fields in Eq.~(57) of Ref.~\cite{Pereira:2025su3}, although
the two fields are now not orthogonal.
In BW gauge, $\Phi_1$ and $\Phi_2$ oscillate as $e^{\mp i\omega t}$ and $\Phi_3$ is static.

We now substitute \eqref{phiansatz} in \eqref{scalarcurrent}. Denoting $\phi_{1\pm}=\eta_1x_\pm$,
$\phi_{10}=\eta_1x_0$, $\phi_{3+}=\eta_3y_+$ and $\phi_{30}=\eta_3y_0$, we find that the
non-vanishing components of the scalar field four-current are
\ba
J^1_t &=& J^6_t = g\,C_t\cos\nu t, \quad J^2_t = J^7_t = g\,C_t\sin\nu t, \nn\\
J^3_t &=& \frac{1}{\sqrt{3}}J^8_t = g\big[(\omega+\nu)\phi_{1+}^2 - (\omega-\nu)\phi_{1-}^2 + \nu\phi_{3+}^2\big], \nn\\
J^1_z &=& -J^6_z = g\,C_z\sin\nu t, \quad J^2_z = -J^7_z = -g\,C_z\cos\nu t, \nn\\
J^4_z &=& 2g\,D_z\sin 2\nu t, \quad J^5_z = -2g\,D_z\cos 2\nu t,
\label{scalarcurrents}
\ea
where
\ba
C_t &=& 2\phi_{10}\Big[\Big(\omega+\frac{\nu}{2}\Big)\phi_{1+} + \Big(\omega-\frac{\nu}{2}\Big)\phi_{1-}\Big] + \nu\phi_{3+}\phi_{30}, \nn\\
C_z &=& f_1\big[(\phi_{1+}+\phi_{1-})^2 + 2\phi_{10}^2 + \phi_{30}^2\big] \nn\\
&& - f_2\big[\phi_{10}(\phi_{1+}-\phi_{1-}) + \phi_{3+}\phi_{30}\big], \nn\\
D_z &=& f_2\big(\phi_{1+}^2+\phi_{1-}^2+\phi_{3+}^2\big) \nn\\
&& - f_1\big[\phi_{10}(\phi_{1+}-\phi_{1-}) + \phi_{3+}\phi_{30}\big].
\ea
The scalar four-current has exactly the same time dependence as the gauge field four-current in
Eqs.~\eqref{gaugecurrent1}--\eqref{gaugecurrent3}, and in particular the ratio
$J^8_t/J^3_t=\sqrt{3}$ is guaranteed by the reflection symmetry that relates $\Phi_1$ and $\Phi_2$.
Matching the $a=3,8$ components of the time component of the four-currents gives
\ba
0 &=& \nu\big(2f_1^2+4f_2^2\big) \nn\\
&& + g^2\big[(\nu+\omega)\phi_{1+}^2 + (\nu-\omega)\phi_{1-}^2 + \nu\phi_{3+}^2\big] ,
\label{alggauss}
\ea
the $a=1,2,6,7$ components of the time component give
\ba
3\nu f_1f_2 &=& g^2\phi_{10}\Big[\Big(\omega+\frac{\nu}{2}\Big)\phi_{1+} + \Big(\omega-\frac{\nu}{2}\Big)\phi_{1-}\Big] \nn\\
&& + \frac{g^2}{2}\nu\phi_{3+}\phi_{30} ,
\label{alggauss2}
\ea
and the $z$ components give
\ba
\nu^2 f_1 &=& \frac{g^2}{2}f_1\big[(\phi_{1+}+\phi_{1-})^2 + 2\phi_{10}^2 + \phi_{30}^2\big] \nn\\
&& - \frac{g^2}{2}f_2\big[\phi_{10}(\phi_{1+}-\phi_{1-}) + \phi_{3+}\phi_{30}\big] ,
\label{algz1}
\\
4\nu^2 f_2 &=& g^2f_2\big(\phi_{1+}^2+\phi_{1-}^2+\phi_{3+}^2\big) \nn\\
&& - g^2f_1\big[\phi_{10}(\phi_{1+}-\phi_{1-}) + \phi_{3+}\phi_{30}\big] .
\label{algz2}
\ea
Equation \eqref{alggauss} is the condition of color neutrality along $H$. The gauge fields carry
a color charge density that is proportional to their electric energy density, and it has to be
canceled by the scalar field components with $\omega+\nu<0$. For the 3-type field the analogous
condition fixes $\eta=\Omega/g$ (Eq.~(53) of Ref.~\cite{Pereira:2025su3}).

Thus far we have demanded that the scalar field four-current matches the gauge field four-current.
Next, we need to demand that the scalar fields satisfy their own equations of motion
\eqref{Phieq}. Writing $\Phi_1 = \eta_1e^{-i\nu\Lambda t}e^{-i\omega t}x$ with
$x=(x_+,x_0,x_-)^T$, and using \eqref{Kmatrix} in \eqref{covDcovDPhi}, we obtain
\be
D_{\mu}D^{\mu}\Phi_1 = -\eta_1e^{-i\nu\Lambda t}e^{-i\omega t}\big[(\omega+\nu\Lambda)^2 - \calK^2\big]x .
\ee
We also have $|\Phi_1|^2=|\Phi_2|^2=\eta_1^2$ and $|\Phi_3|^2=\eta_3^2$. Hence \eqref{Phieq}
becomes the eigenvalue equation $[(\omega+\nu\Lambda)^2-\calK^2]x = M_1^2x$, where we have defined
\be
M_i^2 \equiv m_i^2 + 2\kappa_i\eta_i^2 ,
\label{masses}
\ee
with $\eta_2=\eta_1$. The equation for $\Phi_2$ is equivalent by the reflection symmetry, and
$\Phi_3$ gives a similar equation with $\omega\to0$. Using
\be
\calK^2 = \begin{pmatrix}
f_1^2+f_2^2 & -f_1f_2 & f_1^2\\
-f_1f_2 & 2f_1^2 & f_1f_2\\
f_1^2 & f_1f_2 & f_1^2+f_2^2
\end{pmatrix} ,
\ee
the scalar field equations of motion become
\ba
M_1^2\phi_{1+} &=& \big[(\omega+\nu)^2-f_1^2-f_2^2\big]\phi_{1+} \nn\\
&& - f_1^2\phi_{1-} + f_1f_2\phi_{10}, \label{algphi1}\\
M_1^2\phi_{10} &=& \big(\omega^2-2f_1^2\big)\phi_{10} + f_1f_2(\phi_{1+}-\phi_{1-}), \\
M_1^2\phi_{1-} &=& \big[(\omega-\nu)^2-f_1^2-f_2^2\big]\phi_{1-} \nn\\
&& - f_1^2\phi_{1+} - f_1f_2\phi_{10}, \\
M_3^2\phi_{3+} &=& \big(\nu^2-f_2^2\big)\phi_{3+} + f_1f_2\phi_{30}, \\
M_3^2\phi_{30} &=& -2f_1^2\phi_{30} + 2f_1f_2\phi_{3+} .
\label{algphi3}
\ea

Equations \eqref{alggauss}--\eqref{algphi3} are nine algebraic conditions for the frequency
$\omega$, the amplitudes $\eta_1$ and $\eta_3$, the unit vectors $x$ and $y$, and the mass
parameters $M_1^2$ and $M_3^2$. These are ten numbers subject to the two normalization conditions
\eqref{normalization}. One combination of the conditions is redundant because of current
conservation, which leaves ten independent conditions for ten unknowns. Hence the solutions are
isolated for each value of $\epsilon$. Numerical solutions for several values of $\epsilon$ are
given in Table~\ref{tab:uniform}.

\begin{table*}
\caption{Homogeneous 8-type solutions, Eqs.~\eqref{gaugeField} and \eqref{phiansatz}, for several
values of $\epsilon=gE/\Omega$. Equations \eqref{alggauss}--\eqref{algphi3} are satisfied to
$2\times10^{-14}$.}
\label{tab:uniform}
\begin{ruledtabular}
\begin{tabular}{cccccccccccc}
$\epsilon/\Omega$ & $gW^8_{tz}/\Omega^2$ & $\omega/\Omega$ & $g^2\eta_1^2/\Omega^2$ & $g^2\eta_3^2/\Omega^2$ & $M_1^2/\Omega^2$ & $M_3^2/\Omega^2$ & $x_+$ & $x_0$ & $x_-$ & $y_+$ & $y_0$\\
\hline
1/3 & 0.19245 & $-1.071162$ & 0.503404 & 1.076929 & 0.310676 & 0.247194 & 0.99995 & $-0.00756$ & 0.00574 & 0.70674 & 0.03209\\
1/2 & 0.28868 & $-1.162932$ & 0.506208 & 1.179720 & 0.404214 & 0.244342 & 0.99981 & $-0.01537$ & 0.01185 & 0.70560 & 0.06536\\
1 & 0.57735 & $-1.687375$ & 0.511247 & 1.802103 & 1.265194 & 0.235702 & 0.99870 & $-0.03949$ & 0.03218 & 0.69692 & 0.16910\\
3/2 & 0.86603 & $-2.599064$ & 0.512319 & 2.927802 & 4.075081 & 0.230291 & 0.99743 & $-0.05452$ & 0.04649 & 0.68726 & 0.23527\\
2 & 1.15470 & $-3.894264$ & 0.512407 & 4.549306 & 10.927093 & 0.227368 & 0.99654 & $-0.06248$ & 0.05481 & 0.68065 & 0.27100\\
\end{tabular}
\end{ruledtabular}
\end{table*}

The field $\Phi_1$ lies almost entirely in its first component, $x_+\simeq1$, so in temporal gauge
it rotates with the frequency $\omega+\Omega/2$. The mass parameter of the third field is known in
closed form,
\be
M_3^2 = \frac{1}{8}\bigg[\Omega^2-\epsilon^2+\sqrt{(\Omega^2-\epsilon^2)^2+\frac{32}{9}\Omega^2\epsilon^2}\,\bigg] .
\label{M3}
\ee
At $\epsilon=\Omega$, for example, we find $M_3^2=\Omega^2/(3\sqrt{2})$, $y_+=(1+\sqrt{2})/(2\sqrt{3})$
and $y_0=(\sqrt{2}-1)/\sqrt{6}$. High precision values of the other parameters at $\epsilon=\Omega$
are given in Appendix~\ref{appD}.

As in Ref.~\cite{Pereira:2025su3} (see Eq.~(66) there), the scalar field equations only fix the
combinations \eqref{masses} and not $m_i^2$ and $\kappa_i$ separately. Any $\kappa_i\ge0$ is allowed
with $m_i^2 = M_i^2-2\kappa_i\eta_i^2$. For $\kappa_i=0$ all the $m_i^2$ are positive and
$\Phi_i=0$ is a stable vacuum. Since the three scalar fields are not mutually orthogonal, the
interaction terms $(\Phi_i^\dagger\Phi_j)^2$ have to be absent from the potential. For the
orthogonal 3-type scalar fields these terms vanished automatically.

\subsection{Two scalar fields and other 8-type backgrounds}
\label{twoscalar}

Is the third scalar field really necessary? The answer can be obtained by counting (see
Appendix~\ref{appA}). For a uniform BW gauge field, the gauge field equations of motion give five
independent conditions on the scalar fields, and each scalar field that oscillates at a single
frequency in BW gauge brings two parameters, its frequency and its amplitude. Three scalar fields
therefore generically suffice for given $(\theta,\epsilon)$ and leave a one-parameter family of
solutions, while two scalar fields can only suffice on curves in the $(\theta,\epsilon)$ plane.

\begin{figure}
\subfloat[]{\includegraphics[width=0.45\textwidth,angle=0]{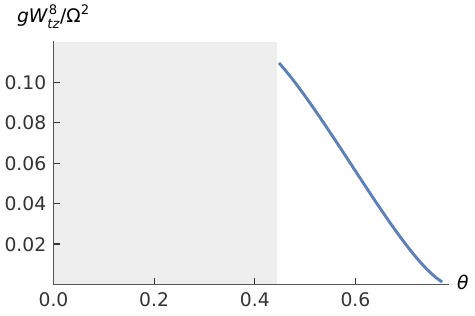}}\\
\subfloat[]{\includegraphics[width=0.45\textwidth,angle=0]{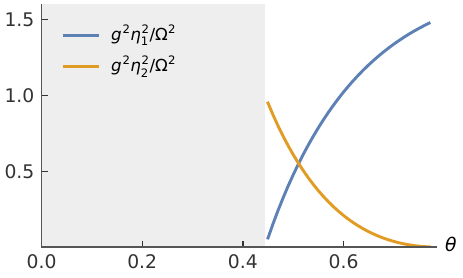}}
\caption{(a) The 8-type electric field $gW^8_{tz}/\Omega^2$ and (b) the scalar field amplitudes
$g^2\eta_i^2/\Omega^2$ vs $\theta$ along the two-scalar branch. No two-scalar solution was found
in the shaded region $\theta<\theta_*\simeq0.444$.}
\label{TwoScalar}
\end{figure}

On the slice $\theta=0$ extensive numerical searches did not find any solution with two scalar
fields, for any value of $\epsilon$. For $\theta\neq0$, where the three frequencies are all
different, we find one branch of two-scalar solutions, which exists for $\theta_*<\theta<\pi/4$
with $\theta_*\simeq0.44402$. It is shown in Table~\ref{tab:two} and Fig.~\ref{TwoScalar}. Along the
branch the two scalar fields are far from orthogonal, $|v_1\cdot v_2|\approx0.9$--$0.98$, where
$v_{1,2}$ are their unit color vectors in BW gauge, and so the interaction term
$\kappa_{12}$ of Ref.~\cite{Pereira:2025su3} must vanish. Equal quartic couplings together with
equal masses would require $\kappa<0$. Hence the two scalar fields can source the 8-type electric
field only on this curve, with $\kappa_{12}=0$ and with $m_1\neq m_2$ or $\kappa_1\neq\kappa_2$.
As $\theta\to\theta_*$, one of the scalar fields becomes infinitely heavy and its frequency
diverges, while $\eta_1^2\omega_1$ stays finite and tends to $-0.3396\,\Omega^3/g^2$. Away from the
slice, three scalar fields sufficed at every point that we tested,
$\theta\in\{0.1,0.2,0.3,0.4\}$ and $\epsilon/\Omega\in\{1/6,1/3,1\}$.

\begin{table*}
\caption{The two-scalar branch of homogeneous 8-type solutions. The scalar fields oscillate at
single frequencies $\omega_{1,2}$ in BW gauge (Appendix~\ref{appA}), and $v_{1,2}$ are their unit
color vectors.}
\label{tab:two}
\begin{ruledtabular}
\begin{tabular}{cccccccccc}
$\theta$ & $\epsilon/\Omega$ & $gW^8_{tz}/\Omega^2$ & $\omega_1/\Omega$ & $\omega_2/\Omega$ & $g^2\eta_1^2/\Omega^2$ & $g^2\eta_2^2/\Omega^2$ & $M_1^2/\Omega^2$ & $M_2^2/\Omega^2$ & $|v_1\cdot v_2|$\\
\hline
0.46 & 0.30367 & 0.10621 & $-2.50530$ & $-0.22044$ & 0.15432 & 0.87132 & 3.76658 & 0.11396 & 0.975 \\
0.50 & 0.29881 & 0.09321 & $-1.02032$ & $-0.21147$ & 0.47646 & 0.60968 & 0.20345 & 0.12312 & 0.969 \\
0.55 & 0.28686 & 0.07512 & $-0.74317$ & $-0.20007$ & 0.78706 & 0.37185 & 0.02856 & 0.13469 & 0.961 \\
0.60 & 0.26853 & 0.05618 & $-0.64956$ & $-0.18857$ & 1.02241 & 0.20849 & 0.00513 & 0.14620 & 0.950 \\
0.65 & 0.24200 & 0.03737 & $-0.60731$ & $-0.17696$ & 1.20177 & 0.10131 & 0.00072 & 0.15756 & 0.938 \\
0.70 & 0.20271 & 0.01989 & $-0.58714$ & $-0.16516$ & 1.33875 & 0.03699 & 0.00003 & 0.16876 & 0.924 \\
\end{tabular}
\end{ruledtabular}
\end{table*}

\subsection{Electric string solution}
\label{Estringsoln}

In the previous subsections, we derived the uniform 8-type electric field solution for all
equations of motion. We now build on the uniform solution to construct electric string (flux tube)
solutions. For the 3-type electric string in Ref.~\cite{Pereira:2025su3}, we attached a single
profile function $f(r)$ to the gauge fields and profile functions $h_i(r)$ to the scalar fields,
where $r$ is the cylindrical radial coordinate. The gauge field equations of motion then gave
$h\propto f$ and a single equation for $f$, Eq.~(78) of Ref.~\cite{Pereira:2025su3}. Two features
of the 3-type solution made this possible. The gauge field rotates at a single frequency
$\Omega$, and the color charge densities of the scalar and gauge fields cancel at every point once
$h\propto f$. Neither feature survives for the 8-type electric string.

First, the 8-type gauge field has components that rotate at the frequency $\Omega/2$ and at the
frequency $\Omega$. Away from the string axis these obey different radial equations, roughly
$\nabla^2f_1+\Omega^2f_1/4\simeq0$ and $\nabla^2f_2+\Omega^2f_2\simeq0$ where
$\nabla^2 = d^2/dr^2+r^{-1}d/dr$, and so they need independent profile functions $f_1(r)$ and
$f_2(r)$.

Second, the string can only be net color neutral. To see this, suppose that the rotation
frequency is the same at every $r$, so that the gauge fields in temporal gauge stay bounded in
time. Then the component of the Gauss law along $H$, which is the analog of
Eq.~\eqref{alggauss}, contains no second derivatives. It becomes a constraint that has to hold
at every $r$, \itie the color charge density along $H$ has to vanish at every point. This
constraint is not implied by the other field equations, because the Bianchi identity involves the
Gauss law only through $[H,\,\cdot\,]$ (see Appendix~\ref{appB}). The radial equations are then
overdetermined by one function. The 3-type string evades this through $h\propto f$, but with two
independent gauge profile functions there is no such structure and generically no solution.
We therefore allow the local rotation frequency $\nu(r)$ to vary slightly with $r$. The Gauss law
along $H$ then becomes an equation for $\nu(r)$, and we only impose net color neutrality,
\itie $\nu(r)\to$ const as $r\to\infty$.

Accordingly, we write the gauge fields in temporal gauge as
\ba
W^{1}_{\mu} &=& \frac{1}{g}\sin\nu t\,\big[2f_1\partial_\mu z + \sqrt{2}\psi'\partial_\mu r\big], \nn\\
W^{2}_{\mu} &=& -\frac{1}{g}\cos\nu t\,\big[2f_1\partial_\mu z + \sqrt{2}\psi'\partial_\mu r\big], \nn\\
W^{6}_{\mu} &=& -\frac{1}{g}\sin\nu t\,\big[2f_1\partial_\mu z - \sqrt{2}\psi'\partial_\mu r\big], \nn\\
W^{7}_{\mu} &=& \frac{1}{g}\cos\nu t\,\big[2f_1\partial_\mu z - \sqrt{2}\psi'\partial_\mu r\big], \nn\\
W^{4}_{\mu} &=& \frac{2}{g}f_2\sin 2\nu t\,\partial_\mu z, \ \
W^{5}_{\mu} = -\frac{2}{g}f_2\cos 2\nu t\,\partial_\mu z, \nn\\
W^{3}_{\mu} &=& -\frac{t}{g}\nu'\partial_\mu r, \ \
W^{8}_{\mu} = -\frac{\sqrt{3}\,t}{g}\nu'\partial_\mu r,
\label{stringW}
\ea
and the scalar fields as
\ba
\Phi_1 &=& \begin{pmatrix}
\phi_{1+} e^{-i(\omega+\nu)t}\\ \phi_{10} e^{-i\omega t}\\ \phi_{1-} e^{-i(\omega-\nu)t}
\end{pmatrix}, \
\Phi_2 = \begin{pmatrix}
-\phi_{1-} e^{i(\omega-\nu)t}\\ \phi_{10} e^{i\omega t}\\ -\phi_{1+} e^{i(\omega+\nu)t}
\end{pmatrix}, \nn\\
\Phi_3 &=& \begin{pmatrix}
\phi_{3+} e^{-i\nu t}\\ \phi_{30}\\ -\phi_{3+} e^{i\nu t}
\end{pmatrix}.
\label{stringphi}
\ea
Here $f_{1,2}$, $\nu$, $\psi$ and $\phi_{1\pm}$, $\phi_{10}$, $\phi_{3+}$, $\phi_{30}$ are functions of
$r$, and primes denote derivatives with respect to $r$. Setting the profile functions to constants
and $\psi'=0$ gives back the homogeneous solution \eqref{Kmatrix}, \eqref{phiansatz}.
Equation \eqref{stringW} is the 8-type counterpart of the 3-type string ansatz in Eq.~(67) of
Ref.~\cite{Pereira:2025su3}, with two profile functions instead of one.

The new ingredients are the radial components of the gauge fields. The angle $\psi(r)$ describes
a slow rotation of the eigenvectors of $H$ with $r$. In the original basis we have
$n=(0,\cos\psi,\sin\psi)^T$ with $\psi(0)=\arctan\sqrt{2}$, and $\psi$ gives bounded radial
components that rotate in time. The components $W^{3,8}_r$ that grow linearly in $t$ describe a
static radial electric field in temporal gauge, since $W^a_{tr}=\partial_tW^a_r$ when $W^a_t=0$,
\be
W^{3}_{tr} = -\frac{\nu'}{g}, \quad W^{8}_{tr} = -\frac{\sqrt{3}\,\nu'}{g} .
\ee
Including the rotating part generated by $\psi'$, the magnitude of the radial electric field is
\be
|W_{tr}| = \frac{2}{g}\sqrt{\nu'^2+\nu^2\psi'^2} ,
\label{Wtr}
\ee
where $2|\nu'|/g$ is the static part along $\Lambda$ and $2\nu|\psi'|/g$ is the rotating part. All
gauge-invariant quantities are time-independent.

Substituting \eqref{stringW} and \eqref{stringphi} in the field equations gives nine second order
equations and one first order constraint, which are listed in Appendix~\ref{appC}. The gauge field
equations are
\ba
\nabla^2\nu &=& \nu\big(\psi'^2+2f_1^2+4f_2^2+g^2\phi_{3+}^2\big) \nn\\
&& + g^2\big[(\nu+\omega)\phi_{1+}^2+(\nu-\omega)\phi_{1-}^2\big],
\label{nuODE}\\
\frac{\Dpsi\nu}{\sqrt{2}} &=& \nu\Big\{3f_1f_2-\frac{g^2}{2}\big[\phi_{10}(\phi_{1+}-\phi_{1-})+\phi_{3+}\phi_{30}\big]\Big\} \nn\\
&& - g^2\omega\,\phi_{10}(\phi_{1+}+\phi_{1-}),
\label{psiODE}\\
\nabla^2f_1 &=& \big(\psi'^2-\nu^2\big)f_1-\frac{\Dpsi f_2}{\sqrt{2}} \nn\\
&& +\frac{g^2}{2}f_1\big[(\phi_{1+}+\phi_{1-})^2+2\phi_{10}^2+\phi_{30}^2\big] \nn\\
&& -\frac{g^2}{2}f_2\big[\phi_{10}(\phi_{1+}-\phi_{1-})+\phi_{3+}\phi_{30}\big],
\label{f1ODE}\\
\nabla^2f_2 &=& \big(\psi'^2-4\nu^2\big)f_2+\sqrt{2}\,\Dpsi f_1 \nn\\
&& +g^2f_2\big(\phi_{1+}^2+\phi_{1-}^2+\phi_{3+}^2\big) \nn\\
&& -g^2f_1\big[\phi_{10}(\phi_{1+}-\phi_{1-})+\phi_{3+}\phi_{30}\big],
\label{f2ODE}
\ea
where $\Dpsi q\equiv\psi''q+\psi'q/r+2\psi'q'$, so that $\Dpsi\nu=(r\nu^2\psi')'/(r\nu)$. The scalar field equations have the same form as
Eqs.~\eqref{algphi1}--\eqref{algphi3}, with additional $\nabla^2$ and $\psi'$ terms. Note that in
\eqref{f1ODE} and \eqref{f2ODE} the $\Omega^2$ of the 3-type equation (78) of
Ref.~\cite{Pereira:2025su3}, $f''+f'/r+\Omega^2f=\ldots$, appears as $\nu^2=\Omega^2/4$ for $f_1$ and
as $4\nu^2=\Omega^2$ for $f_2$. Far from the axis we therefore expect $f_1\sim J_0(\Omega r/2)$ and $f_2\sim J_0(\Omega r)$,
and their ratio cannot stay fixed. Also note that \eqref{nuODE} is the Gauss law along $H$. If
$\nu$ were constant, its right-hand side would have to vanish at every $r$.

We choose boundary conditions such that all profile functions are regular at $r=0$ with
vanishing derivatives. Following Ref.~\cite{Pereira:2025su3}, we take the values on the axis from
the homogeneous solution, here the one with $\epsilon=\Omega/3$ in the first row of
Table~\ref{tab:uniform}, which supplies the unit vectors $x$ and $y$, the frequency $\omega$ and the
mass parameters $M_i^2$. As in Ref.~\cite{Pereira:2025su3} we rescale the homogeneous solution, but
now with independent factors $s$ for the gauge fields and $s_\phi$ for the scalar fields. (For the
3-type string $h\propto f$ ties the two together.) Hence
\ba
f_1(0) &=& s\,\frac{\epsilon}{3}, \quad f_2(0) = s\,\frac{\epsilon}{6}, \quad \nu(0) = \frac{\Omega}{2}, \nn\\
\phi_{1\pm}(0) &=& s_\phi\eta_1x_\pm, \quad \phi_{10}(0) = s_\phi\eta_1x_0, \nn\\
\phi_{3+}(0) &=& s_\phi\eta_3y_+, \quad \phi_{30}(0) = s_\phi\eta_3y_0,
\label{bcs}
\ea
and the values on the axis are not themselves a homogeneous solution. Since $f_2(0)=f_1(0)/2$, the
electric field on the axis is exactly of 8-type. The angle $\psi$ only enters through its
derivatives, and so $\psi(0)$ is a gauge choice. For the parameters of the model we take
$\kappa_i=0.25$ and $m_i^2=M_i^2-2\kappa_i\eta_i^2$, which gives
\ba
m_1^2 &=& m_2^2 = 0.058974\,\Omega^2, \quad m_3^2 = -0.291270\,\Omega^2, \nn\\
\omega &=& -1.071162\,\Omega .
\label{params}
\ea
With these values every scalar field component oscillates at large $r$, \itie
$(\omega_i+\mu_j)^2>m_i^2$, where $\omega_{1,2}=\pm\omega$ and $\omega_3=0$ are the frequencies of the
scalar fields in BW gauge and $\mu_j$ are the eigenvalues of $H$. Hence there are no modes that grow with $r$ and need to be suppressed.
However, since $m_3^2<0$, $\Phi_3=0$ is tachyonic. Homogeneous fluctuations of $\Phi_3$ grow in time
at a rate up to $\sqrt{|m_3^2|}\simeq0.54\,\Omega$, and so the asymptotic state of this string is
classically unstable, as for the 3-type strings with $m^2<0$ in Ref.~\cite{Pereira:2025su3}.

The only remaining condition is net color neutrality, $\nu(r)\to$ const. For a generic
value of $s_\phi/s$, $\nu(r)$ drifts linearly at large $r$, because the oscillating tails of the
gauge and scalar fields carry a net charge. The sign of the drift changes as $s_\phi/s$ crosses
$0.40$. We fix $s_\phi/s$ by requiring zero mean drift over a window $\Omega r\in[L/2,L]$, and then
increase $L$. Table~\ref{tab:roots} in Appendix~\ref{appD} shows that the tuned ratio converges, to
$s_\phi/s=0.4049013$ for $s=0.6$. This plays the role of the condition $F_0\le1$ in
Ref.~\cite{Pereira:2025su3}, except that here a parameter has to be tuned rather than bounded.

\begin{figure}
\subfloat[]{\includegraphics[width=0.45\textwidth,angle=0]{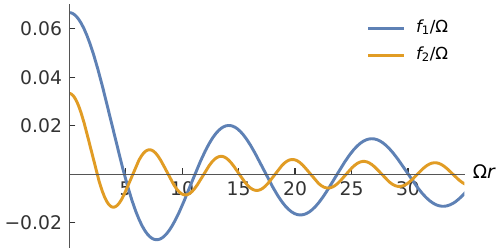}}\\
\subfloat[]{\includegraphics[width=0.45\textwidth,angle=0]{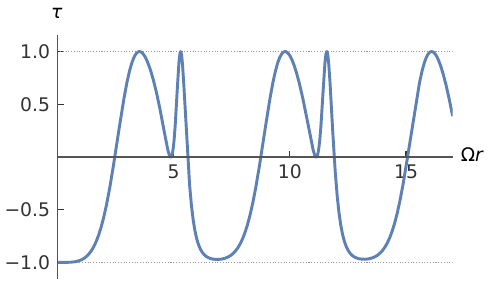}}
\caption{(a) Gauge field profile functions $f_1$ (frequency $\nu\simeq\Omega/2$) and $f_2$
(frequency $2\nu$) and (b) the type ratio $\tau$ of Eq.~\eqref{tauf} vs $\Omega r$ for $s=0.6$,
$\kappa_i=0.25$ and $g=1$. The electric field is of 8-type ($\tau=-1$) on the axis.}
\label{GaugeProfiles}
\end{figure}

\begin{figure}
\includegraphics[width=0.45\textwidth,angle=0]{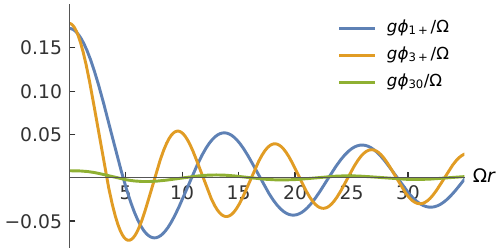}
\caption{The dominant scalar field profile functions vs $\Omega r$ for $s=0.6$, $\kappa_i=0.25$
and $g=1$.}
\label{ScalarProfiles}
\end{figure}

\begin{figure}
\subfloat[]{\includegraphics[width=0.45\textwidth,angle=0]{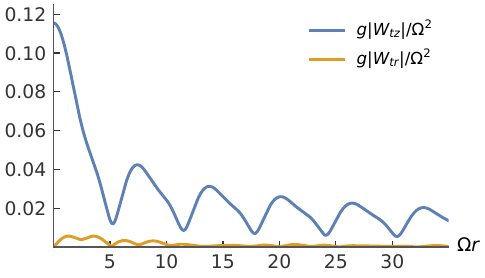}}\\
\subfloat[]{\includegraphics[width=0.45\textwidth,angle=0]{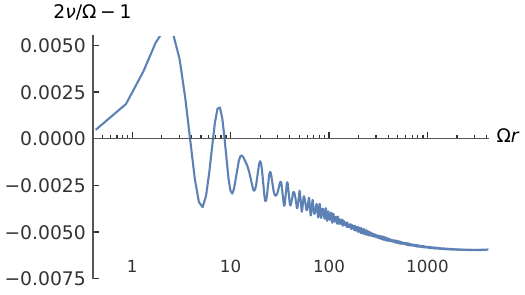}}
\caption{(a) Magnitudes of the axial electric field $|W_{tz}|$ and of the radial electric field
$|W_{tr}|$ and (b) the relative variation of the local rotation frequency $\nu(r)$ vs
$\Omega r$ for $s=0.6$. The frequency converges at large $r$, and so the string is net color
neutral.}
\label{RadialField}
\end{figure}

\begin{figure}
\includegraphics[width=0.45\textwidth,angle=0]{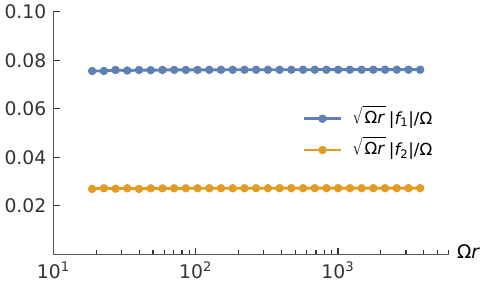}
\caption{Maxima of $\sqrt{\Omega r}\,|f_{1,2}|/\Omega$ in logarithmic bins of $\Omega r$. The
constant envelopes show that the profile functions decay as $r^{-1/2}$.}
\label{Envelopes}
\end{figure}

The solution for $s=0.6$ is shown in Figs.~\ref{GaugeProfiles}--\ref{Envelopes}. On the axis
$f_2/f_1=1/2$, so that $\tau=-1$ and $|W_{tz}|=0.1155\,\Omega^2/g$. The gauge profile functions
behave like $J_0(\Omega r/2)$ and $J_0(\Omega r)$ with nonlinear corrections. Their first zeros
are at $\Omega r=4.88$ and $2.47$, compared with $4.81$ and $2.40$ for the Bessel functions.
Consequently $\tau$ departs from $-1$ away from the axis. It crosses $0$ at $\Omega r=2.47$,
where $f_2$ vanishes, reaches $+1$ (an 8-type field of opposite orientation) at $\Omega r=3.53$,
and then oscillates. The string therefore has an 8-type core rather than an 8-type electric field
everywhere. Since the two components rotate at different frequencies, and therefore have different
radial wavenumbers, we expect this to be generic.

The frequency $\nu(r)$ changes by at most $0.6\%$ and converges to $0.49703\,\Omega$. The static
radial electric field $2|\nu'|/g$ peaks at $0.0054\,\Omega^2/g$ near $\Omega r=3.6$, which is
$4.7\%$ of the axial electric field on the axis, and including the rotating part
$|W_{tr}|\le0.0057\,\Omega^2/g$. At large $r$, $r\nu'$ oscillates with a constant amplitude of
about $0.018\,\Omega$ and zero mean. The radial electric field therefore falls off as $1/r$, and
the enclosed color charge oscillates about zero without secular growth. This is the precise sense
in which the string is net color neutral.

The gauge and scalar profile functions decay as $r^{-1/2}$, with constant envelopes out to
$\Omega r\approx4\times10^3$ (Fig.~\ref{Envelopes}), while $\nu$ approaches a constant and $\nu'$
and $\psi'$ fall off as $1/r$. The energy density therefore falls off as $1/r$ and the energy per
unit length grows linearly with the transverse size. By our reading of its large-$R$ behavior,
the same is true for the $J_0$-type tails of the 3-type profile function in Eq.~(78) of
Ref.~\cite{Pereira:2025su3}.

\subsection{Summary of electric string solution}
\label{summary}

Finally, we summarize the electric string solution:
\ba
W^{1}_{\mu} &=& \frac{1}{g}\sin\nu t\,\big[2f_1\partial_\mu z + \sqrt{2}\psi'\partial_\mu r\big], \nn\\
W^{2}_{\mu} &=& -\frac{1}{g}\cos\nu t\,\big[2f_1\partial_\mu z + \sqrt{2}\psi'\partial_\mu r\big], \nn\\
W^{6}_{\mu} &=& -\frac{1}{g}\sin\nu t\,\big[2f_1\partial_\mu z - \sqrt{2}\psi'\partial_\mu r\big], \nn\\
W^{7}_{\mu} &=& \frac{1}{g}\cos\nu t\,\big[2f_1\partial_\mu z - \sqrt{2}\psi'\partial_\mu r\big], \nn\\
W^{4}_{\mu} &=& \frac{2}{g}f_2\sin 2\nu t\,\partial_\mu z, \ \
W^{5}_{\mu} = -\frac{2}{g}f_2\cos 2\nu t\,\partial_\mu z, \nn\\
W^{3}_{\mu} &=& -\frac{t}{g}\nu'\partial_\mu r, \ \
W^{8}_{\mu} = -\frac{\sqrt{3}\,t}{g}\nu'\partial_\mu r,
\ea
with the scalar fields $\Phi_1$, $\Phi_2$ and $\Phi_3$ in \eqref{stringphi}. The model parameters
are given in \eqref{params}, with $\kappa_i=0.25$ and no interaction terms between the scalar
fields. The profile functions solve the radial equations of Appendix~\ref{appC} with the boundary
conditions \eqref{bcs}, where $\epsilon=\Omega/3$, $s=0.6$ and $s_\phi/s=0.4049013$. They are
provided as ancillary files with this paper. The electric field is of 8-type on the axis, with
$|W_{tz}|=0.1155\,\Omega^2/g$. The profile functions $f_{1,2}$ and $\phi$ decay as $r^{-1/2}$.
There is a weak static radial electric field, at most $4.7\%$ of the axial electric field, that
falls off as $1/r$, and the enclosed color charge oscillates about zero. Tuned solutions also
exist for $s=0.4$. For $s=0.8$--$1$ the drift of $\nu$ changes sign at moderate $r$, but we have
not established convergence at large $r$.

\section{Schwinger pair production}
\label{unexciting}

In this section, we consider quantum fluctuations about the 8-type electric field. As in
Ref.~\cite{Pereira:2025su3}, the motivation is to show that the electric field is stable to
Schwinger pair production of gauge bosons which, in temporal gauge, is equivalent to particle
production in a time-dependent background~\cite{Vachaspati:2022ktr}.

We denote the background gauge fields in \eqref{gaugeField} by $A^a_\mu$ and consider
fluctuations $Q^a_\mu$ in temporal gauge, $W^a_\mu = A^a_\mu + Q^a_\mu$. As in
Ref.~\cite{Pereira:2025su3}, we combine the fields as
\ba
w^{\pm}_{\mu} &=& W^2_{\mu} \pm iW^1_{\mu}, \quad
y^{\pm}_{\mu} = W^5_{\mu} \pm iW^4_{\mu}, \nn\\
z^{\pm}_{\mu} &=& W^7_{\mu} \pm iW^6_{\mu},
\ea
and write
\ba
w^{\pm}_{\mu} &=& e^{\mp i\nu t}\Big[-\frac{2f_1}{g}\partial_\mu z + q^{\pm}_{\mu}\Big], \nn\\
z^{\pm}_{\mu} &=& e^{\mp i\nu t}\Big[\frac{2f_1}{g}\partial_\mu z + p^{\pm}_{\mu}\Big], \nn\\
y^{\pm}_{\mu} &=& e^{\mp 2i\nu t}\Big[-\frac{2f_2}{g}\partial_\mu z + s^{\pm}_{\mu}\Big], \nn\\
W^3_\mu &=& Q^3_\mu, \quad W^8_\mu = Q^8_\mu,
\label{flucdef}
\ea
where $\nu=\Omega/2$, $f_1=\epsilon/3$ and $f_2=\epsilon/6$ as in \eqref{Kmatrix}. Compared to
the 3-type case in Ref.~\cite{Pereira:2025su3}, all three SU(2) subgroups now carry a background
field, and the phase factors are $e^{\mp i\nu t}$, $e^{\mp i\nu t}$ and $e^{\mp 2i\nu t}$.

The phase factors in \eqref{flucdef} are those of the time-dependent gauge transformation
$V(t)=e^{-i\nu\Lambda t}$ that appears in \eqref{Kmatrix}. In matrix form, \eqref{flucdef} reads
\be
\frac{g}{2}W^a_\mu\lambda^a = V(t)\big[\calK\,\partial_\mu z + q_\mu\big]V(t)^\dagger ,
\ee
where the matrix $q_\mu=(g/2)q^a_\mu\lambda^a$ collects $q^\pm_\mu$, $p^\pm_\mu$, $s^\pm_\mu$,
$Q^3_\mu$ and $Q^8_\mu$. Since $V(t)$ is a gauge transformation, the gauge fields are gauge
equivalent to
\be
\calA'_\mu = \nu\Lambda\,\partial_\mu t + \calK\,\partial_\mu z + q_\mu ,
\label{rotframe}
\ee
in which the background is independent of time. The Lagrangian \eqref{lag} is gauge invariant,
and so after substituting \eqref{flucdef} it contains no explicit time dependence. The time
derivatives of the phase factors combine with the fluctuations into covariant derivatives with the
constant matrix $\nu\Lambda$, for example $\partial_t q^\pm_\mu \to (\partial_t\mp i\nu)q^\pm_\mu$
and $\partial_t s^\pm_\mu \to (\partial_t\mp 2i\nu)s^\pm_\mu$, in the same way as in
Ref.~\cite{Pereira:2025su3}.

The same holds for the scalar fields. In the gauge \eqref{rotframe}, the background scalar
fields \eqref{phiansatz} become $V^\dagger\Phi_i=e^{-i\omega_it}c_i$ with constant vectors $c_i$
and $\omega_1=\omega$, $\omega_2=-\omega$, $\omega_3=0$. We write the scalar field fluctuations
as
\be
\Phi_i + \delta\Phi_i = V(t)\,e^{-i\omega_it}\big(c_i+\xi_i\big) .
\label{scalarfluc}
\ee
Each scalar field oscillates at a single frequency, and the potential \eqref{potentialV} contains
no interaction terms between different scalar fields. The factor $e^{-i\omega_it}$ therefore only
shifts the time derivative of $\xi_i$ by the constant $-i\omega_i$, and the Lagrangian for the
fluctuations $\xi_i$, including their couplings to the gauge field fluctuations, has no explicit
time dependence either. This would not be the case with the interaction terms
$\kappa_{ij}[(\Phi_i^\dagger\Phi_j)^2+{\rm h.c.}]$ of Ref.~\cite{Pereira:2025su3}, which would bring
in the factors $e^{\pm2i(\omega_i-\omega_j)t}$. Unlike in Ref.~\cite{Pereira:2025su3}, we therefore
do not need to assume that the scalar fields are heavy.

Thus, the dynamics of all the fluctuations about the homogeneous 8-type electric field take place
in a time-independent background, and, as for the 3-type electric field, they will not be excited.
Equivalently, their quanta will not be produced. The two different frequencies $\nu$ and $2\nu$ do
not affect this argument, since both phase factors in \eqref{flucdef} are generated by the single
matrix $\Lambda$. The argument does not rely on our choice $\theta=0$ either, since for any BW gauge
field the time dependence in temporal gauge is generated in the same way (see also
Appendix~\ref{appD}). As for the 3-type electric field, this argument does not
address classical stability, \itie whether the time-independent equations for the fluctuations have
growing modes.

For the electric string, the same construction applies with $V(t)$ replaced by the
$r$-dependent gauge transformation that relates the temporal gauge fields \eqref{stringW} and
\eqref{stringphi} to the static form $\calA_t=H(r)$, $\calA_z=iK(r)$,
$\Phi_i=e^{-i\omega_it}\phi_i(r)$ of Appendix~\ref{appC}. The components of $W^{3,8}_r$ that grow
linearly in $t$ are generated by the $r$ dependence of $\nu(r)$ in this transformation. The
background for the fluctuations is then again independent of time, which is the criterion used for
the 3-type electric string in Ref.~\cite{Pereira:2025su3}.

However, the 8-type string differs from the 3-type string in one respect. For the 3-type string,
$\calA_t=H$ is the same at every $r$, whereas for the 8-type string $H(r)$ depends on $r$. This is
the static radial electric field of Sec.~\ref{Estringsoln}. For a static but inhomogeneous
background, the absence of explicit time dependence does not by itself rule out pair production.
For example, the Maxwellian electric field $W^a_\mu=Ez\,\partial_\mu t\,\delta^{ab}$ is also
independent of time, and it produces pairs by tunneling. For the string of Sec.~\ref{Estringsoln},
$\nu(r)$ lies between $0.4970\,\Omega$ and $0.5030\,\Omega$, so the gauge field fluctuations that
are charged under $H$ see potential differences of at most about $0.006\,\Omega$ and
$0.012\,\Omega$, while far from the string they are massless. Whether the radial electric field
leads to pair production therefore requires an analysis of the spectrum of fluctuations about the
string, which we leave for future work.

\section{Conclusions}
\label{conclusions}

We have constructed 8-type electric field and electric string solutions in SU(3) gauge theory with
scalar fields in the fundamental representation, as summarized in Sec.~\ref{summary}. This
answers the question that was left open in Ref.~\cite{Pereira:2025su3}. The construction follows
the procedure in Refs.~\cite{Vachaspati:2022gco,Vachaspati:2023tpt,Pereira:2025su3}, but differs
from the 3-type case in several ways. All of them can be traced back to the fact that in temporal
gauge the 8-type gauge field rotates within three SU(2) subgroups with at least two different
frequencies, whereas the 3-type gauge field rotates within a single SU(2) subgroup at one
frequency. As a result, the homogeneous 8-type electric field generically needs three scalar
fields rather than two, the electric string needs two gauge profile functions and is of 8-type
on its axis but not everywhere, and the string is only net color neutral, so that it carries
a weak radial electric field that is strongest in the core. As for the 3-type electric field, the
homogeneous 8-type electric field is stable to Schwinger pair production of gauge bosons, since all
the time dependence of the fluctuations can be removed by a time-dependent field redefinition
(Sec.~\ref{unexciting}).

Our 8-type string has $m_3^2<0$, as the 3-type strings with $m^2<0$ in
Ref.~\cite{Pereira:2025su3}, and so its asymptotic state is classically unstable. A string with
all $m_i^2>0$ requires the suppression of one mode that grows with $r$, which amounts to one more
shooting condition. In addition, the oscillatory $r^{-1/2}$ tails make the energy per unit length
grow with the transverse size, which, on our reading, is also the case for the 3-type string.

There are a number of follow-up questions that we plan to address in the future.
Does the static radial electric field of the 8-type string lead to pair production of gauge
bosons (see Sec.~\ref{unexciting})? Are the 8-type solutions classically stable, as in the
SU(2) case~\cite{Pereira:2024sic}? Can an 8-type electric string be constructed on the two-scalar
branch of Sec.~\ref{twoscalar}, \itie with the field content of Ref.~\cite{Pereira:2025su3}? It will be
interesting to see if such solutions could also be studied using lattice methods and whether
fundamental scalar fields could arise as emergent degrees of freedom in pure non-Abelian gauge theory.

\acknowledgements
I am grateful to Tanmay Vachaspati for motivation and discussions.
This work was supported by the U.S. Department of Energy, Office of High Energy Physics, under Award No.~DE-SC0019470.

\appendix
\section{Sourcing uniform electric fields with single-frequency scalar fields}
\label{appA}

In this Appendix we describe the general method behind Secs.~\ref{type8scalar}
and~\ref{twoscalar}. In BW gauge $\calA_t=H$ and $\calA_z=B$ are constant, and so a stationary
configuration needs time-independent four-currents. We take each scalar field to oscillate at a
single frequency, $\Phi_i=\eta_ie^{-i\omega_it}v_i$ with $|v_i|=1$. The scalar field equation of
motion then becomes the eigenvalue problem
\be
\big[(\omega_i+H)^2-B^2\big]v_i = M_i^2v_i ,
\ee
with $M_i^2$ as in \eqref{masses}, and the gauge field equations of motion become
\ba
\frac{g^2}{2}\sum_i\eta_i^2\{\omega_i+H,P_i\}\tl &=& [B,[H,B]], \nn\\
\frac{g^2}{2}\sum_i\eta_i^2\{B,P_i\}\tl &=& [H,[H,B]],
\label{appmatch}
\ea
where $P_i=v_iv_i^\dagger$ and $X\tl$ denotes the traceless part of $X$. Here we have used the SU(3)
identity
\be
\sum_a(\lambda^a)_{ij}(\lambda^a)_{kl} = 2\delta_{il}\delta_{jk}-\frac{2}{3}\delta_{ij}\delta_{kl} .
\ee

Since $H$ is real symmetric and $B$ is imaginary antisymmetric, the $v_i$ can be chosen to be
real. Their four-currents then automatically lie in the components $a=1,3,4,6,8$ for the time
component and $a=2,5,7$ for the $z$ component, as required by the BW currents. Covariant
conservation of the four-current of each mode, which is equivalent to
$[(\omega_i+H)^2-B^2,P_i]=0$, removes three of these eight equations and leaves five. This gives
the counting in Sec.~\ref{twoscalar}.

Whether any number of single-frequency scalar fields can source a given BW gauge field is a convex
question, since the required four-current has to lie in the cone generated by the mode currents
for all frequencies and eigenvectors. We solved this linear program on an $8\times7$ grid with
$\theta\in\{0,\dots,0.7\}$ and $\epsilon/\Omega\in\{0.03,\dots,2.7\}$. It is feasible at 53 points.
At the 3 remaining points, all at the smallest value of $\epsilon$, the result is inconclusive.

For the 3-type gauge fields, $v_\pm=(1,\pm1,0)^T/\sqrt{2}$ with $\eta=\Omega/g$ and
$\omega_\pm=\mp(\Omega/2+\epsilon^2/2\Omega)$ reproduce the solution of Ref.~\cite{Pereira:2025su3},
with $M^2=(\epsilon^2/4)(\epsilon^2/\Omega^2-1)$. For the 8-type gauge fields with $\theta=0$, the
reflection $\sigma$ of Sec.~\ref{type8scalar} pairs a mode with frequency $\omega$ with a mode with
frequency $-\omega$, and allows a third, static mode. Equations \eqref{appmatch} then reduce to
\eqref{alggauss}--\eqref{algphi3}.

\section{Frequencies and type of the electric field}
\label{appB}

In the eigenbasis of $H$, the electric field $F=-i[H,B]$ has the matrix elements
$F_{jk}=-i(\mu_j-\mu_k)B_{jk}$. Its diagonal elements vanish and
\ba
\tr F^3 &=& 6\,{\rm Re}(F_{12}F_{23}F_{31}) \nn\\
&=& -6(\mu_1-\mu_2)(\mu_2-\mu_3)(\mu_3-\mu_1) \nn\\
&& \times{\rm Im}(B_{12}B_{23}B_{31}) .
\ea
Suppose that one off-diagonal element of $F$ vanishes. Then the characteristic polynomial of $F$
is $-\lambda^3+\lambda\sum|F_{jk}|^2$, the eigenvalues are $0$ and $\pm(\sum|F_{jk}|^2)^{1/2}$,
and $F$ is of 3-type. An 8-type electric field therefore needs all $F_{jk}\neq0$, \itie distinct
$\mu_j$ and all $B_{jk}\neq0$. Since $[H,[H,B]]_{jk}=(\mu_j-\mu_k)^2B_{jk}$, and $|\mu_3-\mu_1|$ is
larger than both of the other differences if $\mu_2$ lies between $\mu_1$ and $\mu_3$, at least two
different frequencies appear. Conversely, if $[H,[H,B]]=\chi^2B$, at most two of the $B_{jk}$ are
non-vanishing and $F$ is of 3-type.

For the gauge field \eqref{Kmatrix}, the electric field has the matrix elements $F_{+0}=\nu f_1$,
$F_{0-}=-\nu f_1$ and $F_{+-}=2\nu f_2$ in the basis $(u_+,u_0,u_-)$. Hence
$\tr F^2=4\nu^2(f_1^2+2f_2^2)$ and $\tr F^3=-12\nu^3f_1^2f_2$, which gives \eqref{tauf}. The
condition $(\tr F^3)^2=(\tr F^2)^3/6$ is equivalent to $(4x-1)^2(x+2)=0$ with $x=f_2^2/f_1^2$, \itie
to $f_2=\pm|f_1|/2$.

Finally, consider the trace of the Gauss law with $H$ for a stationary configuration in which the
eigenvalues of $H$ do not depend on $r$. Writing $\calA_z=B=iK$ with $K$ real and
antisymmetric, the result is the first order constraint
\be
\tr[H,K]^2+\tr H'^2+g^2\sum_i\phi_i^T H(\omega_i+H)\phi_i = 0 ,
\ee
in which both gauge field terms are non-negative. The Bianchi identity does not contain this component of the Gauss law, which only enters through
$[H,\,\cdot\,]$. Hence it is an independent constraint at every point, which the 3-type solution
satisfies through $h\propto f$ but which the 8-type solution generically cannot satisfy.
Equation \eqref{nuODE} is this constraint once $\nu$ is allowed to vary with $r$.

\section{Radial equations}
\label{appC}

With $\nabla^2q=q''+q'/r$, $\Dpsi q=\psi''q+\psi'q/r+2\psi'q'$ and the local mass parameters
$M_i^2(r)\equiv m_i^2+2\kappa_i|\Phi_i|^2$, where $|\Phi_1|^2=\phi_{1+}^2+\phi_{10}^2+\phi_{1-}^2$ and
$|\Phi_3|^2=2\phi_{3+}^2+\phi_{30}^2$, the scalar field equations are
\ba
\nabla^2\phi_{1+} &=& \big[M_1^2+f_1^2+f_2^2+\psi'^2/2-(\omega+\nu)^2\big]\phi_{1+} \nn\\
&& +\big(f_1^2-\psi'^2/2\big)\phi_{1-} \nn\\
&& -f_1f_2\phi_{10}-\Dpsi\phi_{10}/\sqrt{2}, \\
\nabla^2\phi_{10} &=& \big[M_1^2+2f_1^2+\psi'^2-\omega^2\big]\phi_{10} \nn\\
&& -f_1f_2(\phi_{1+}-\phi_{1-}) \nn\\
&& +\Dpsi(\phi_{1+}-\phi_{1-})/\sqrt{2}, \\
\nabla^2\phi_{1-} &=& \big[M_1^2+f_1^2+f_2^2+\psi'^2/2-(\omega-\nu)^2\big]\phi_{1-} \nn\\
&& +\big(f_1^2-\psi'^2/2\big)\phi_{1+} \nn\\
&& +f_1f_2\phi_{10}+\Dpsi\phi_{10}/\sqrt{2}, \\
\nabla^2\phi_{3+} &=& \big[M_3^2+f_2^2+\psi'^2-\nu^2\big]\phi_{3+} \nn\\
&& -f_1f_2\phi_{30}-\Dpsi\phi_{30}/\sqrt{2}, \\
\nabla^2\phi_{30} &=& \big[M_3^2+2f_1^2+\psi'^2\big]\phi_{30} \nn\\
&& -2f_1f_2\phi_{3+}+\sqrt{2}\,\Dpsi\phi_{3+} .
\ea
Together with \eqref{nuODE}--\eqref{f2ODE} these are the nine radial equations. The $r$ component
of the gauge field equations gives the constraint
\ba
0 &=& f_1'f_2-f_1f_2'+\sqrt{2}\,\psi'\Big(f_1^2+\frac{f_2^2}{2}-\frac{\nu^2}{2}\Big) \nn\\
&& +\frac{g^2}{2}\big[\phi_{10}(\phi_{1+}-\phi_{1-})'-\phi_{10}'(\phi_{1+}-\phi_{1-})\big] \nn\\
&& +\frac{g^2}{2}\big[\phi_{30}\phi_{3+}'-\phi_{3+}\phi_{30}'\big] \nn\\
&& +\frac{g^2\psi'}{2\sqrt{2}}\big[(\phi_{1+}-\phi_{1-})^2+2\phi_{10}^2 \nn\\
&& \hskip 2 cm +2\phi_{3+}^2+\phi_{30}^2\big] ,
\ea
which holds automatically for regular solutions because of the Bianchi identity.

These equations are equivalent to a compact matrix system in BW gauge, with $\calA_t=H(r)$ and
$\calA_z=iK(r)$, where $K$ is real and antisymmetric,
\ba
\nabla^2\phi_i &=& \big[m_i^2+2\kappa_i|\phi_i|^2-(\omega_i+H)^2-K^2\big]\phi_i, \nn\\
\nabla^2H &=& [K,[H,K]]+\frac{g^2}{2}\sum_i\{\omega_i+H,\phi_i\phi_i^T\}\tl, \nn\\
\nabla^2K &=& -[H,[H,K]]+\frac{g^2}{2}\sum_i\{K,\phi_i\phi_i^T\},
\ea
which is the form that we integrated numerically. We checked the component equations above
against the numerical solution. Every equation holds to about $10^{-5}$ of its largest
individual term or better, limited by finite differencing of the tabulated profile functions, and
the constraint holds to $10^{-14}$.

\section{Numerical methods and checks}
\label{appD}

We solved the algebraic equations \eqref{alggauss}--\eqref{algphi3} and \eqref{appmatch} with a
nonlinear least-squares method~\cite{Virtanen:2019joe}. For $\epsilon=\Omega$ we refined the solution with Newton's method
at 60 digits, which gives
\ba
\omega/\Omega &=& -1.6873749884364787947, \nn\\
g^2\eta_1^2/\Omega^2 &=& 0.51124723923839915025, \nn\\
g^2\eta_3^2/\Omega^2 &=& 1.8021028272307736559, \nn\\
M_1^2/\Omega^2 &=& 1.2651940699008742897, \nn\\
x_+ &=& 0.99870186489655161258, \nn\\
x_0 &=& -0.03948753258484501635, \nn\\
x_- &=& 0.03217638610084689034 . \nn
\ea
We then checked the homogeneous solutions against the Euler--Lagrange equations of
\eqref{lag}, derived by computer algebra for all $W^a_t$, all $W^a_z$ and every real scalar field
component. The residuals are below $10^{-14}$, while a 1\% change of one amplitude gives
residuals of order $10^{-2}$. As a direct check of the formulas as written, we also constructed
the temporal gauge fields \eqref{gaugeField} and \eqref{phiansatz} from Table~\ref{tab:uniform}
and evaluated the field equations with exact time derivatives. The residuals are below
$3\times10^{-15}$ for every row. A second implementation, written independently from the
Lagrangian, reproduces these results. Finally, the full Lagrangian \eqref{lag} with the
fluctuations \eqref{flucdef} and \eqref{scalarfluc}, for random values of the fluctuations and their
derivatives, is independent of time to $2\times10^{-15}$. Changing a phase factor, omitting
$e^{-i\omega_it}$, or adding a $\kappa_{12}$ term gives time dependence.

We integrated the radial equations for the string outward from regular data at $r=0$ with the
eighth-order Runge--Kutta method of Dormand and Prince~\cite{Hairer:1993}, as implemented in
SciPy~\cite{Virtanen:2019joe}, with relative tolerance $10^{-10}$--$10^{-12}$. The neutrality
ratio $s_\phi/s$ was tuned by bracketing. Its convergence as the window grows is shown in
Table~\ref{tab:roots}. Each larger window needs more digits, since the neutral configuration is
an unstable equilibrium of the radial evolution.

\begin{table}
\caption{Convergence of the neutrality-tuned ratio $s_\phi/s$ with the size of the window
$\Omega r\in[L/2,L]$. The string solution shown in Figs.~\ref{GaugeProfiles}--\ref{Envelopes} uses
the value for $L=4.16\times10^3$.}
\label{tab:roots}
\begin{ruledtabular}
\begin{tabular}{ccc}
$L$ & $s_\phi/s$ at $s=0.4$ & $s_\phi/s$ at $s=0.6$\\
\hline
$1.04\times10^3$ & 0.4036467669 & 0.4048970330\\
$2.08\times10^3$ & 0.4036530418 & 0.4049009423\\
$4.16\times10^3$ & 0.4036543054 & 0.4049013097\\
$8.31\times10^3$ & 0.4036544429 & 0.4049013256\\
\end{tabular}
\end{ruledtabular}
\end{table}

We constructed the temporal gauge fields \eqref{stringW} and \eqref{stringphi} directly from the
tabulated profile functions and evaluated the full time-dependent gauge and scalar field
equations by finite differences. For $\Omega t\in[0,32]$ and $\Omega r\in[0.7,71]$ the residuals
are below $1.2\times10^{-5}$ of the largest term at each point, whereas rescaling one scalar field
by 1\% raises them by more than four orders of magnitude. A second implementation, which only used
the tabulated profile functions and the formulas above, found residuals of at most
$1.4\times10^{-5}$ relative to the largest term, and typically $10^{-7}$. It also confirmed that
$|W_{tz}|$, $|W_{tr}|$, $|W_{rz}|$ and $|\Phi_i|$ are time-independent to $10^{-11}$, and that at late
times the residuals decrease as the fourth power of the finite-difference step, as they should if
the components of $W^{3,8}_r$ that grow linearly in $t$ are consistent with the field equations.

The profile functions are tabulated for $0<\Omega r\le104$ in steps of $0.0087$, and more coarsely
up to $\Omega r\approx4\times10^3$. They are provided as ancillary files with this paper, together
with scripts that check the solutions.

\bibstyle{aps}
\bibliography{paper}

\end{document}